\documentclass[preprint]{elsarticle}

\journal{High Energy Density Physics}

\usepackage{graphicx}

\begin{document}

\begin{frontmatter}
\title{Increase of the Density, Temperature and Velocity of Plasma Jets driven by a Ring of High Energy Laser Beams}
\author[rice]{Wen Fu}
  \ead{Wen.Fu@rice.edu}
\author[rice]{Edison P. Liang}
\address[rice]{Department of Physics and Astronomy, Rice University, Houston, TX 77005, USA}
\author[chicago]{Milad Fatenejad}
\author[chicago]{Donald Q. Lamb}
\address[chicago]{Flash Center for Computational Science, University of Chicago, Chicago, IL 60637, USA}
\author[michigan]{Michael Grosskopf}
\address[michigan]{Department of Atmospheric, Oceanic and Space Sciences, University of Michigan, Ann Arbor, MI 48109, USA}
\author[llnl]{Hye-Sook Park}
\author[llnl]{Bruce Remington}
\address[llnl]{Lawrence Livermore National Laboratory, Livermore, CA 94550, USA}
\author[princeton]{Anatoly Spitkovsky}
\address[princeton]{Department of Astrophysical Sciences, Princeton University, NJ, 08544, USA}

\begin{abstract}
Supersonic plasma outflows driven by multi-beam, high-energy lasers, such as Omega and NIF, have been and will be used as platforms for a variety of laboratory astrophysics experiments. Here we propose a new way of launching high density and high velocity, plasma jets using multiple intense laser beams in a hollow ring formation. We show that such jets provide a more flexible and versatile platform for future laboratory astrophysics experiments. Using high resolution hydrodynamic simulations, we demonstrate that the collimated jets can achieve much higher density, temperature and velocity when multiple laser beams are focused to form a hollow ring pattern at the target, instead of focused onto a single spot. We carried out simulations with different ring radii and studied their effects on the jet properties. Implications for laboratory collisionless shock experiments are discussed.
\end{abstract}


\end{frontmatter}

\section{Introduction}
Supersonic, well collimated jets are observed from various astrophysical objects, such as protostellar disks \cite{Pech12}, X-ray binaries \cite{Hao09} and active galactic nuclei \cite{Ferrari98}, which exhibit wide range of size and radiation power. Despite their rich phenomenology, many questions on the physics of astrophysical jets, e.g. launching mechanism, collimation mechanism, role of magnetic field and interaction with ambient medium, still remain unanswered. While the traditional approaches to address these questions are mainly direct observation and theoretical modelling, laboratory produced jets with proper scaling relations \cite{Ryutov99, Remington06} may provide an alternative platform to study jets on astrophysical scales. In particular, jets formed by laser produced plasma have been successfully created by either irradiating a planar target with one ``concaved'' (intensity lower at the center) laser beam \cite{Sizyuk07, Pisarczyk07, Tikhonchuk08, Kasperczuk06, Nicolai06, Nicolai08, Kmetik12} or shining multiple beams onto a cone-shaped target \cite{Farley99, Gregory08}. In both designs ablated plasma plumes are produced at different locations on the target and collide on the central vertical axis. The radiative collapse effect, which reduces the pressure along the axis, then channels the plasma flow along the axis and help maintain a jet-like structure \cite{shigemori00}. Here with application to laboratory collisionless shock experiments in mind \textbf{\cite{Ross12, Park12}}, we propose a new design to generate versatile laboratory plasma jets with a large dynamic range. It utilizes multiple intense high-energy laser beams focused onto a simple planar target to form a hollow ring pattern. 

\begin{figure}
\begin{center}
\includegraphics[width=0.7\textwidth]{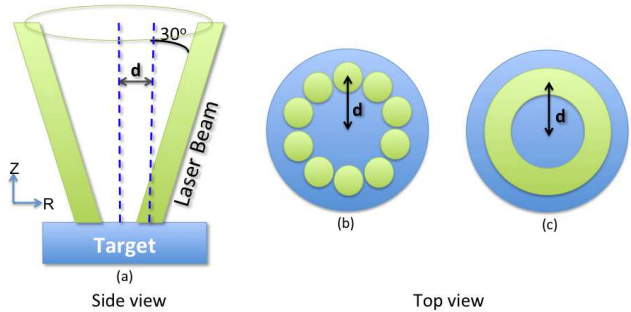}
\caption{\label{fig:design} (Color online) Design of numerical laser-produced plasma jet experiment. (a) side view; (b) top view of target surface (model for ten beams).}
\end{center}
\end{figure}

\section{Numerical Simulation}
The setup of our numerical simulations is illustrated in Fig.~\ref{fig:design}, which is based on the Omega laser parameters used in recent collisionless shock experiments \cite{Ross12}. Here 5 kJ of total laser energy (from ten Omega beams) is used to irradiate a planar plastic (CH) target. Each beam produces a supergaussian focal spot at 351 nm wavelength with a diameter of 250 $\rm{\mu m}$. The target foil measures 2 mm in diameter and 0.5 mm in thickness. These beams hit the target with an incident angle of $30^{\circ}$ from target normal, delivering $\sim 5\,\rm{kJ}$ total energy in a 1 ns square pulse. If all beams hit the same spot at the center of the target, this gives a combined intensity of $\sim 10^{16}\,\rm{W/cm^2}$. We carry out axisymmetric radiative hydrodynamic simulations using the FLASH code \cite{Fryxell00} in a 2D R-Z cylindrical domain. In real experiments, though, the laser energy deposition would not be in a perfectly uniform ring as in Fig.~\ref{fig:design}(b). However, by defocusing the focal spot of each beam or using more beams (e.g.,  use all the 30 beams on one side at Omega), we could come up with a similar pattern (i.e., laser energy covers almost a full circle of the targe surface). In this proof of principle study, we neglect the intensity variation along the ring and defer the study of its effect to future full 3D simulations. The radius of the ring $d$ is the distance between the center of an individual laser beam and the center of the target. The simulation domain is covered by a uniform grid with resolution of $(N_r\times N_z)=(512\times 2560)$ (i.e. $\sim$ 2 $\rm{\mu m}$ resolution) and open boundaries. The domain is initially filled with low density ($2\times 10^{-7}\,\rm{g/cm^3}$) Helium to mimic a vacuum environment. The new FLASH code (version 4.0) recently added the capability to model laser-driven High Energy Density Physics (HEDP) experiments. It solves laser ray tracing in the geometric optics limit and deposits laser energy via inverse-Bremsstrahlung process. The evolution of an unmagnetized plasma is described in a three-temperature fashion (i.e., ions, electrons and radiation are modelled separately). The radiation field is treated by multigroup radiation diffusion and material properties are incorporated with tabulated EOS and opacity data.  

\begin{figure}
\begin{center}
\includegraphics[width=0.9\textwidth]{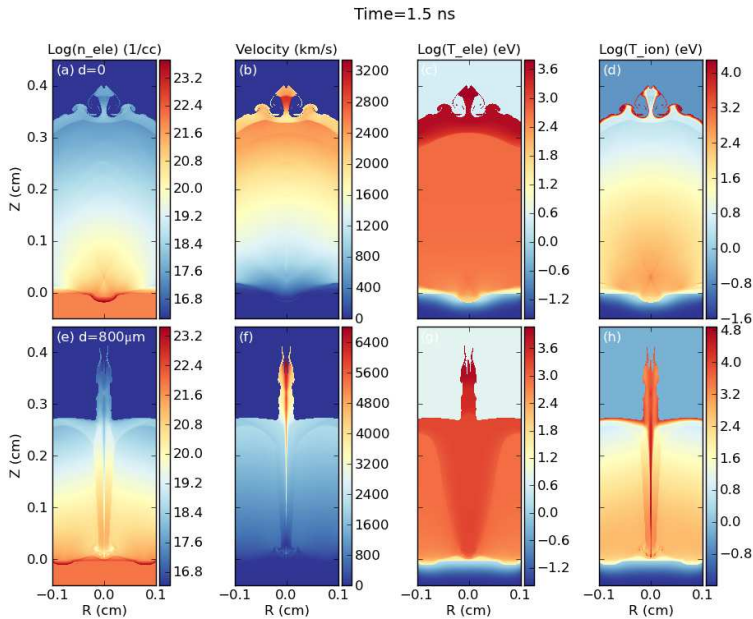}
\caption{\label{fig:compare} (Color online) Snapshots of electron density, flow velocity, electron and ion temperatures at $t=1.5$ ns for two different runs. In one case, all the beams hit the target center (upper panels). In the other case, the focal spot is 800 $\mu$m away from the target center (bottom panels). }
\end{center}
\end{figure}

\begin{figure}
\begin{center}
\includegraphics[width=0.8\textwidth]{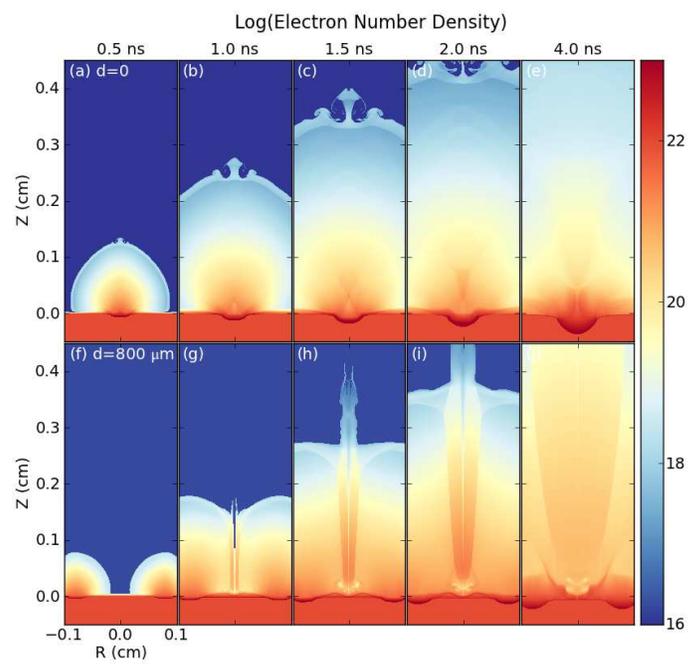}
\caption{\label{fig:evolve} (Color online) Evolution of the plasma jet produced with the radius of on-target beam ring pattern being 0 (upper) and 800 $\rm{\mu m}$ (lower). }
\end{center}
\end{figure}

Fig.~\ref{fig:compare} compares the state of ablated plasma at $t=$ 1.5 ns (0.5 ns after the laser is turned off). In the upper panels, all the laser beams focus on the center of the target. As expected, the laser energy heats up the target and produces a quasi-spherical plasma plume which expands into the ambient low density helium. The shock ``tip'' in the center and two ``roll-up''s on the side seen in all four sub-panels are probably caused by the angle formed between inclined beams and the presence of ambient Helium. Note that in actual experiments, which take place in a vacuum, there is no shear between CH flow and ambient medium, these features would be much less pronounced or even absent. Other than these features, distributions of plasma properties in this case clearly show quasi-spherical uniformity. In contrast, beams in the bottom panels form a ring of 800 $\rm{\mu m}$ radius. And they produce a cylindrical plume  along a ring around the center. As it expands, the plume collides on axis and create a highly collimated outflow with high density, high velocity and high temperature along the axis (see also Fig.~\ref{fig:evolve}). In panels (e), (f) and (h), we can see the formed jet structure is on average about one beam size in width, and its boundary is marked by a sharp transition in plasma conditions. This feature is less prominent in electron temperature plot. Compared with the case with $d=0$, although the laser irradiation intensity is lower (the same total energy gets more spread out), the jet head travels a similar distance ($z\sim0.4 cm$) because of higher velocity along the axis. The formation of the high-density, high-velocity jet is due to the rocket nozzle effect and the long, thin, channel visible on the $z$-axis in panel (e) is caused by the rarefication from the on-axis plasma collision. \textbf{Unlike previous studies (e.g. \cite{Farley99}) with massive targets (Au, Cu, Ag, etc.) where radiative effects play an important role in forming jets, we model a CH target, thus radiative effects should be relatively negligible. This is indeed verified by our runs with radiation being turned off in FLASH. }

Evolution of the plasma jet is depicted in Fig.~\ref{fig:evolve} where the laser beams from a ring of 800 $\rm{\mu m}$ radius.  By $t=$ 4 ns (3 ns past the end of laser energy input), the jet has evolved into a structure as wide as the target. And the jet head has gone way beyond the boundary of our simulation box. \textbf{For the application to collisionless shock experiments, we emphasize that the ``precursor jet'' ahead of the bulk flow has too little mass to really affect the collision dynamics, so that the jet within the bulk flow can still be reliably used to generate collisionless shocks. }

\begin{figure}
\begin{center}
$
\begin{array}{cc}
\includegraphics[width=0.45\textwidth]{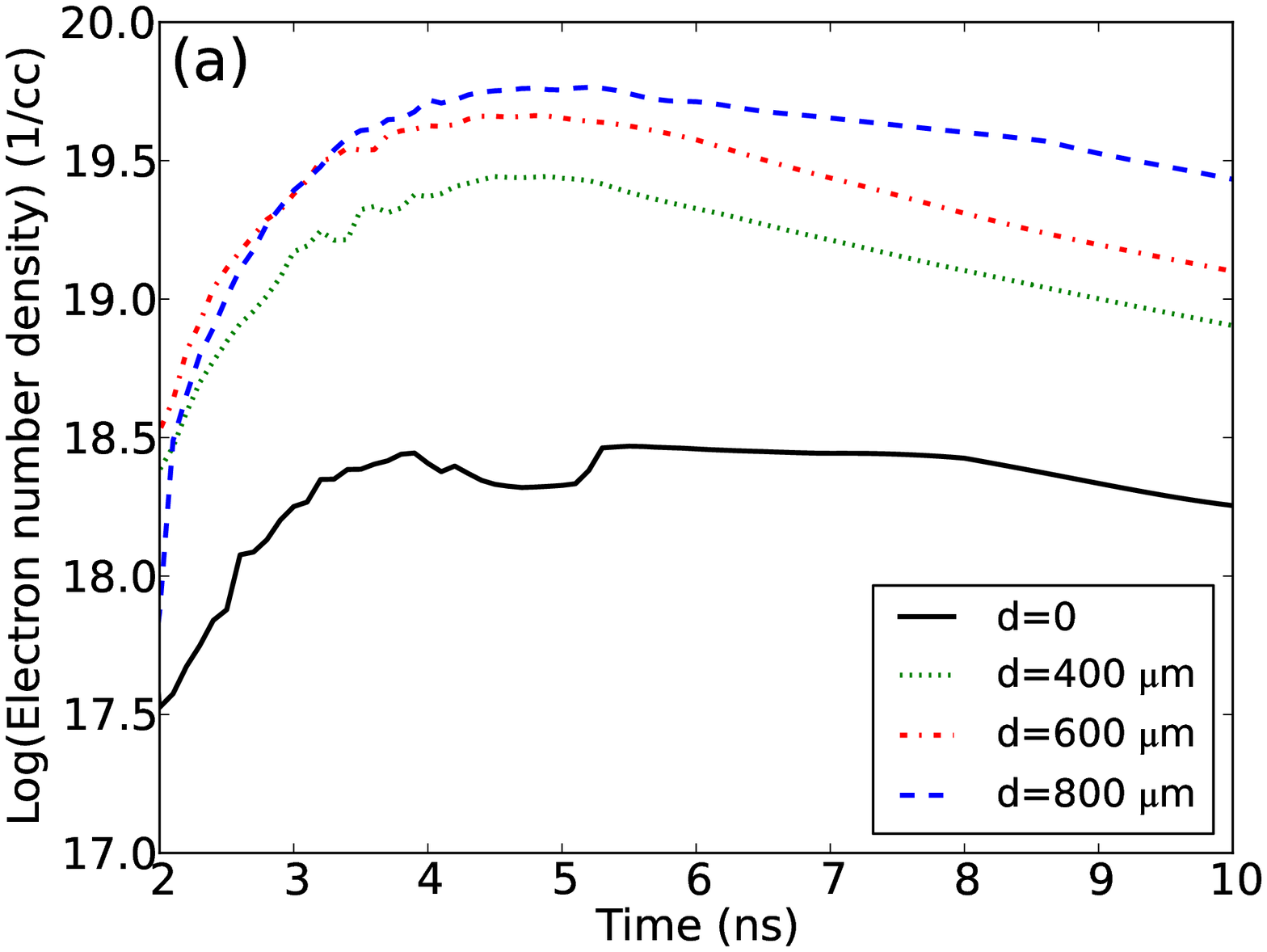} &
\includegraphics[width=0.45\textwidth]{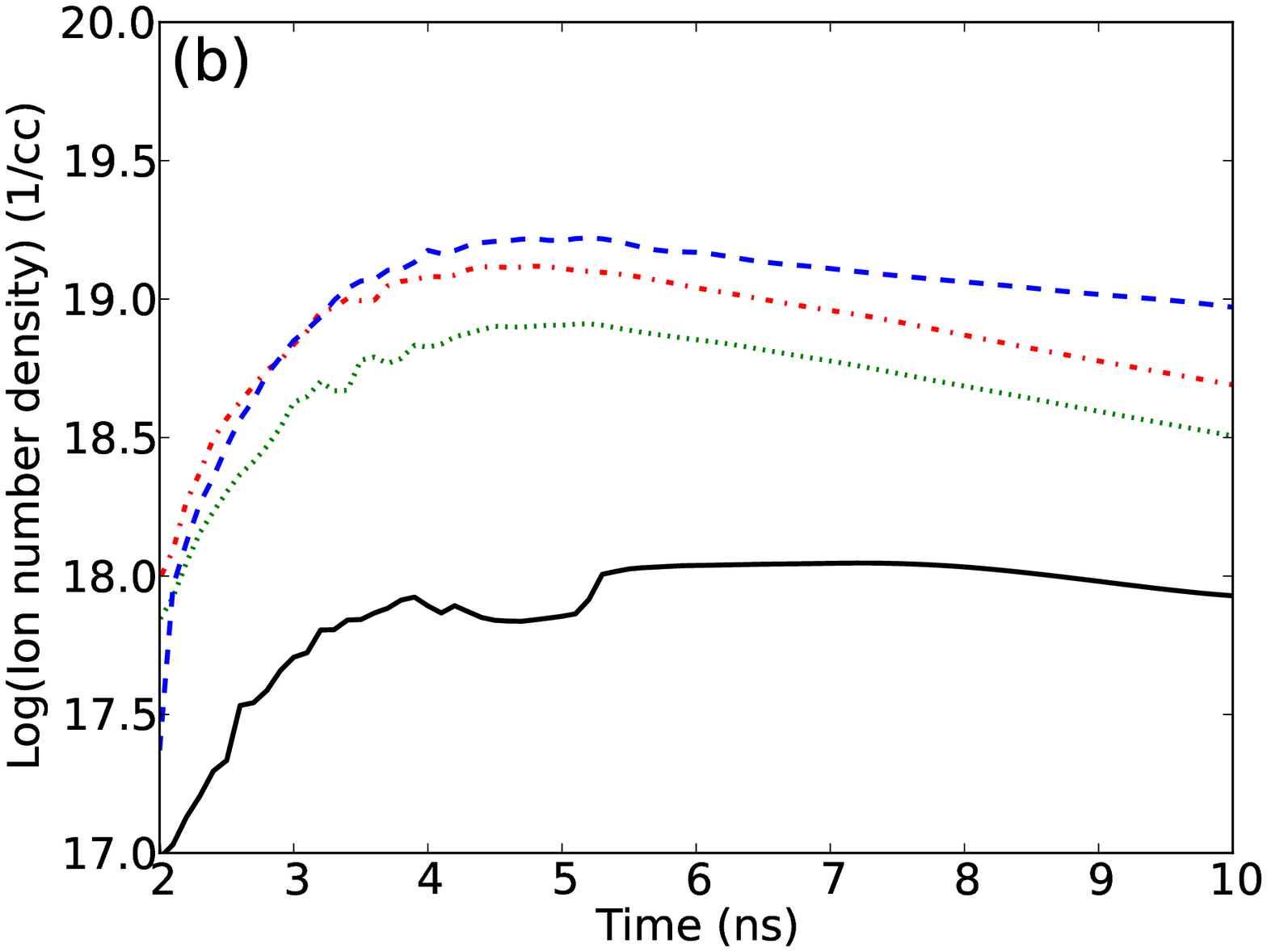} \\
\includegraphics[width=0.45\textwidth]{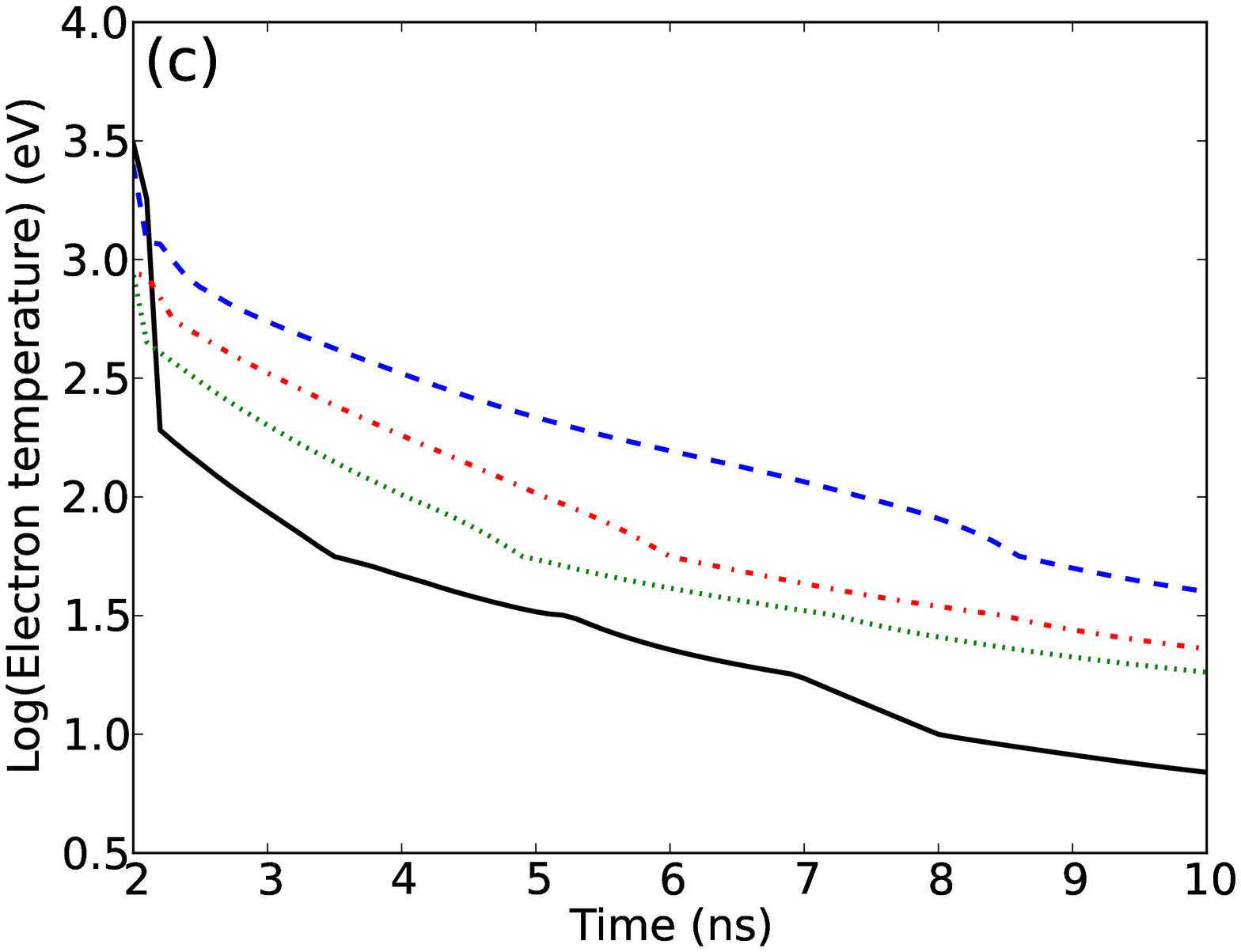} &
\includegraphics[width=0.45\textwidth]{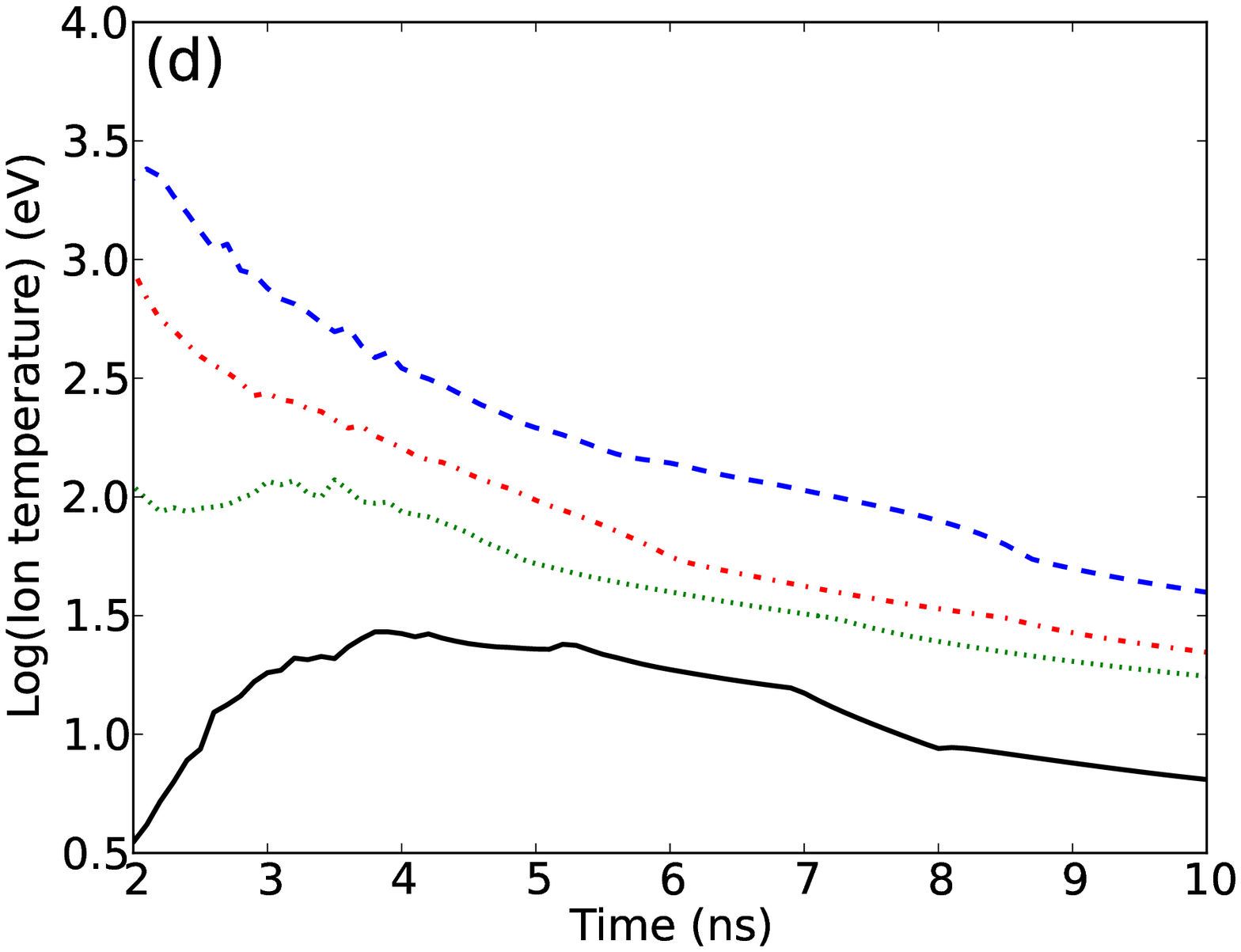} \\
\includegraphics[width=0.45\textwidth]{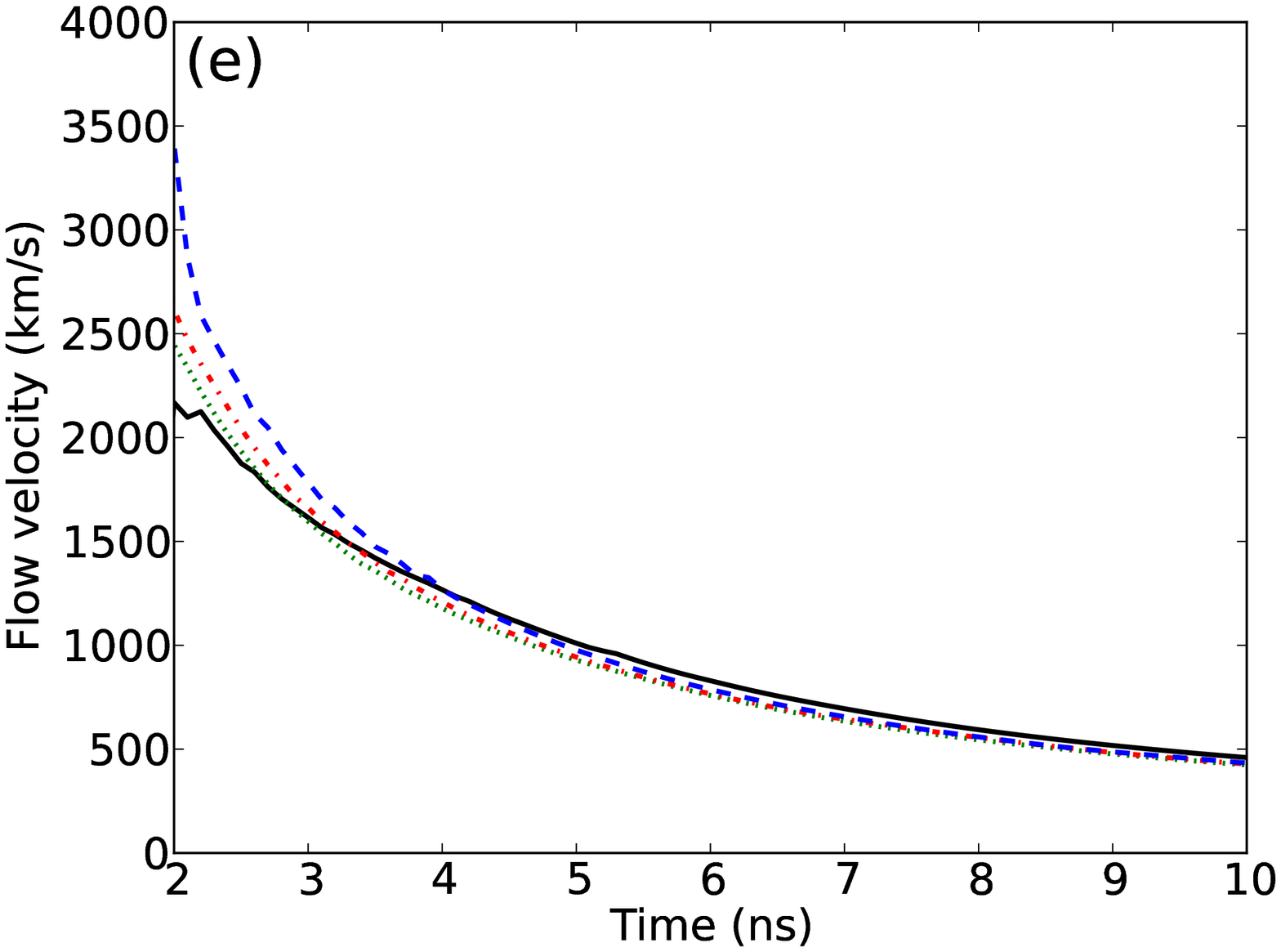} &
\includegraphics[width=0.45\textwidth]{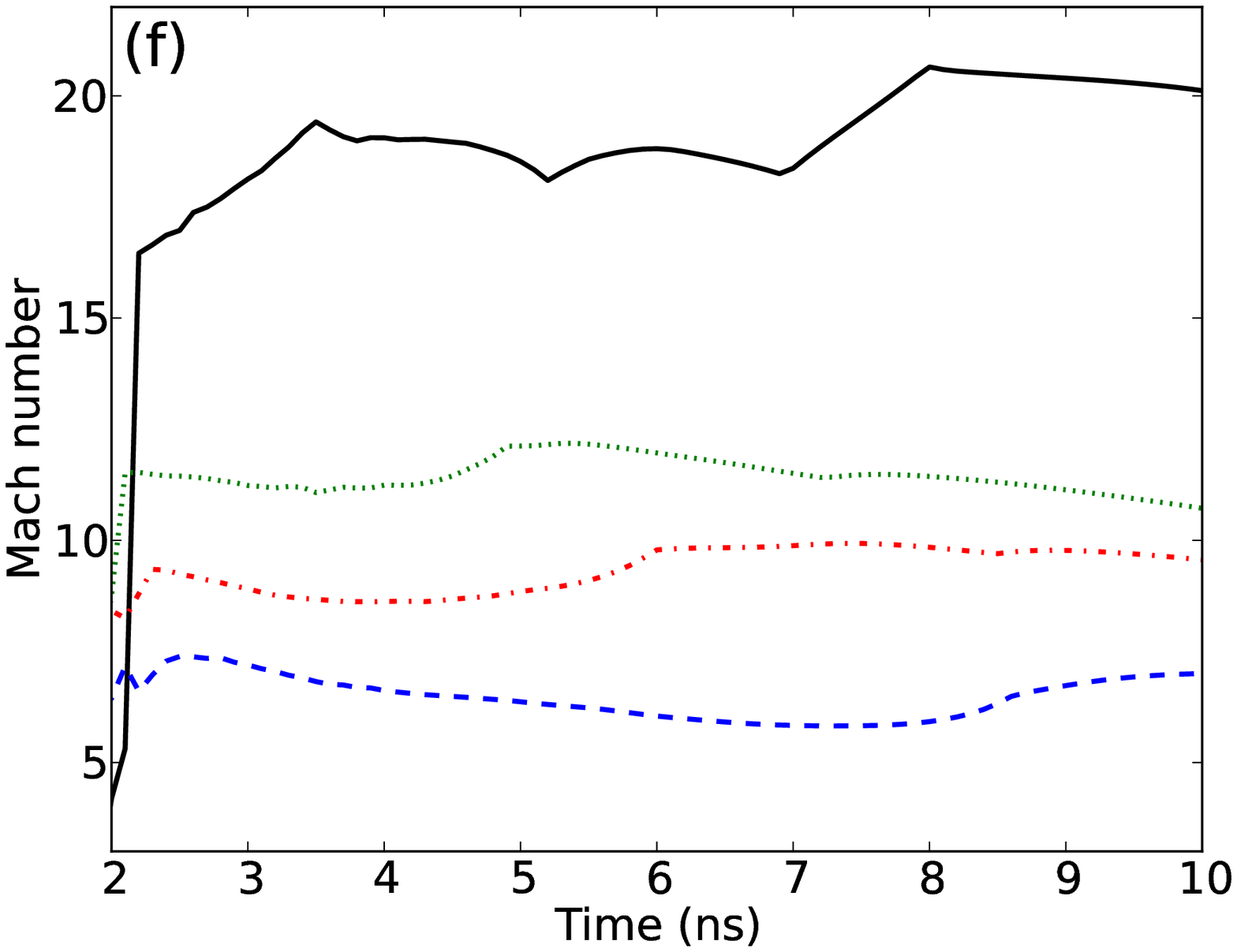} \\
\includegraphics[width=0.45\textwidth]{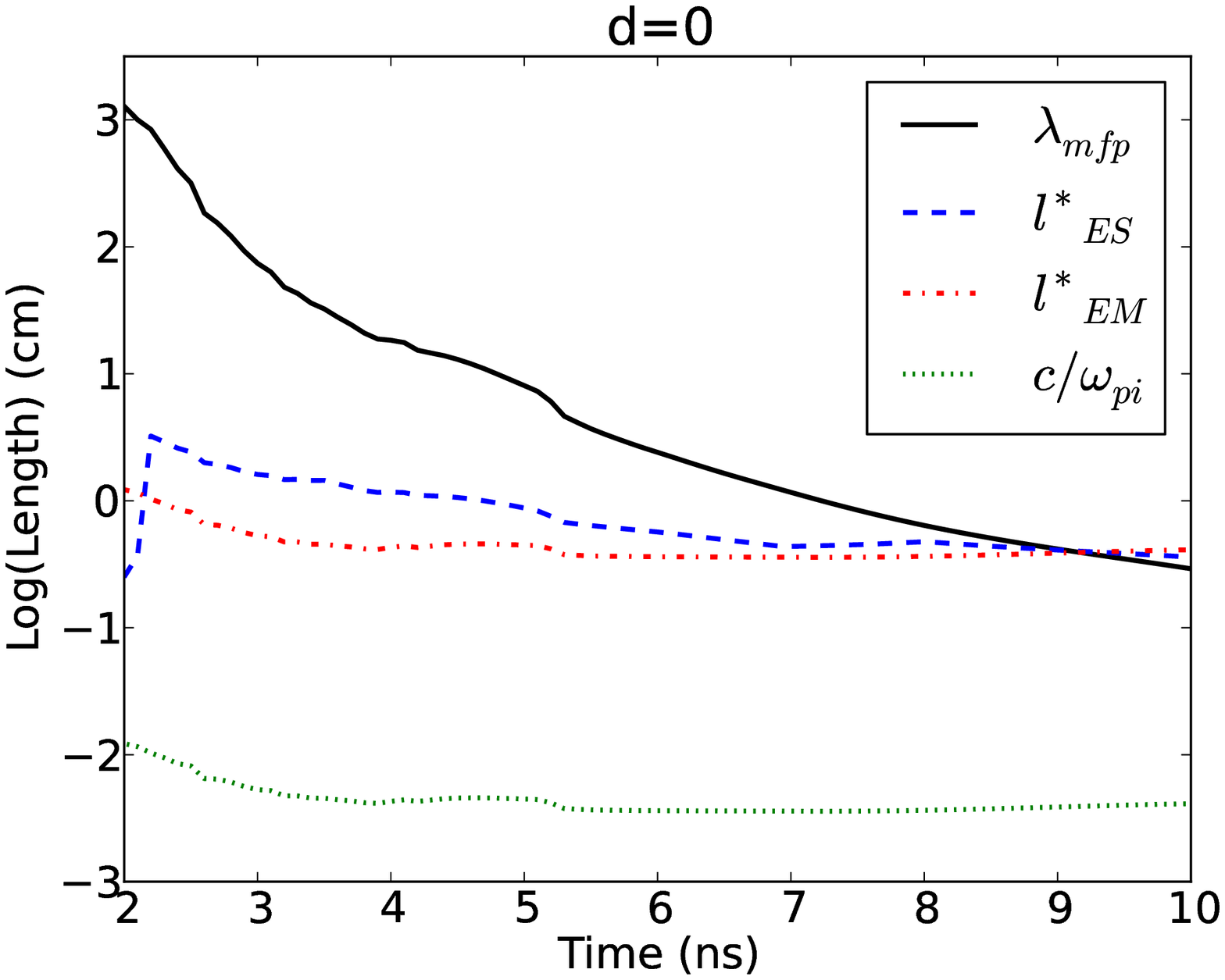} &
\includegraphics[width=0.45\textwidth]{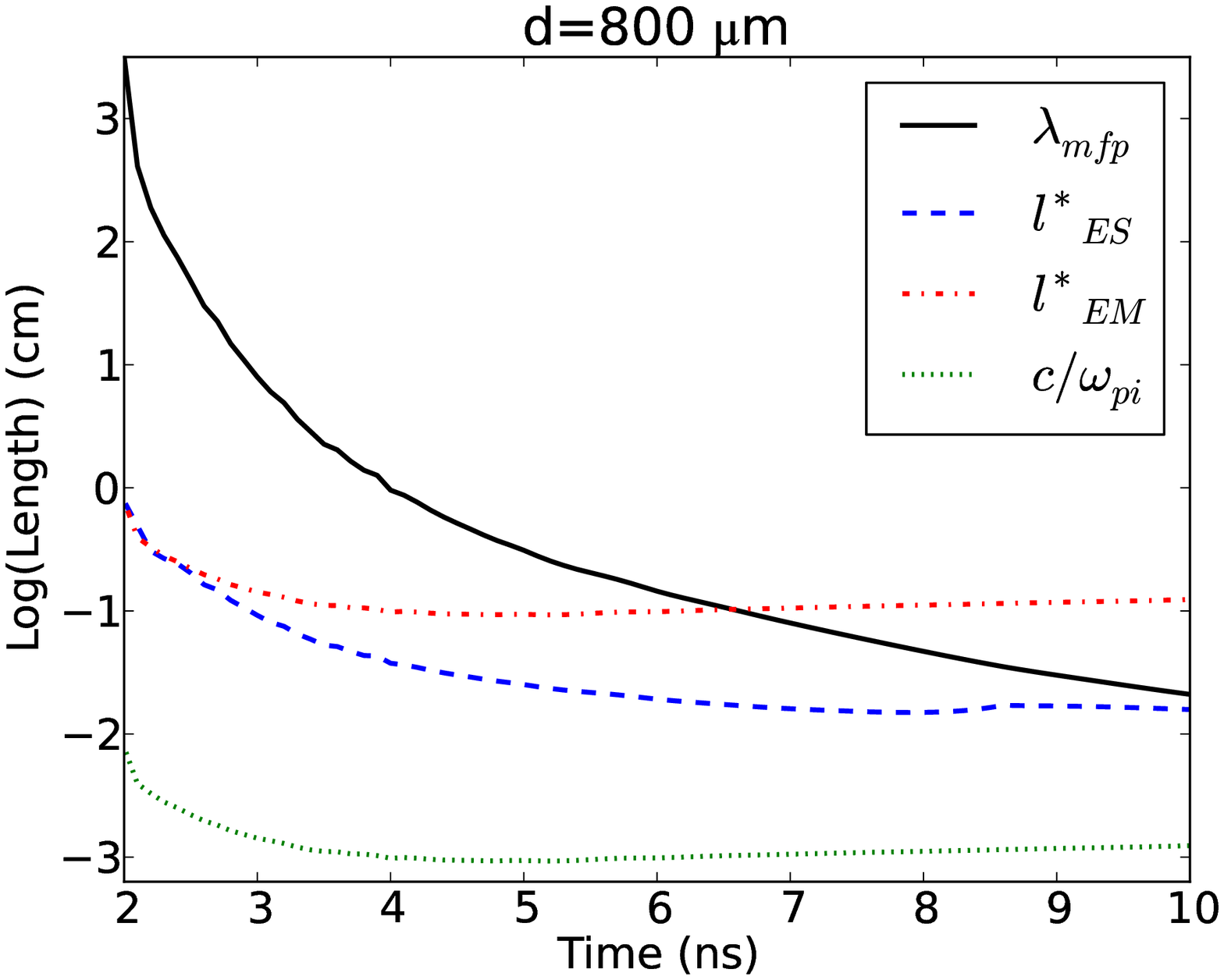} 
\end{array}
$
\caption{\label{fig:fig4} (Color online) Average plasma conditions as a function of time measured at 4 mm above the target surface along the axis. They are the electron, ion number densities (a, b), electron, ion temperatures (c, d), flow velocity and Mach number (e, f). Different line types represents different degrees of beam separations. The bottom two panels show collision mean free path between two counter-steaming flows (black solid), characteristic electrostatic (blue dashed), electromagnetic (red dot-dashed) instability length scales and ion skin depth (green dotted) for two different laser beams separation cases.}
\end{center}
\end{figure}

To study quantitatively the effects of hollow beam radius on the jet properties, we plot in Fig.~\ref{fig:fig4}(a)-(f) the electron, ion number densities, electron, ion temperatures, flow velocity and Mach number as a function of time for four different runs. The Mach number is computed as the ratio of flow velocity to sound speed ($c_s=\sqrt{\gamma P/\rho}$). All these quantities are taken from 4 mm above the target surface and averaged over one beam size (250 $\rm{\mu m}$) around the axis. Fig.~\ref{fig:fig4}(a) and Fig.~\ref{fig:fig4}(b) show that as the laser beams become more separated, both electron and ion number density on the axis increase. All four curves have maxima near $t=5$ ns but from $d=0$ to $d=800\,\rm{\mu m}$ the maximum density goes up by almost one order of magnitude. This is due to the fact that jet collimation becomes more prominent with larger beam separation (plasma more concentrated on the axis). In Fig.~\ref{fig:fig4}(c) and Fig.~\ref{fig:fig4}(d), particle temperatures are predominantly higher, although the increase is less apparent than in the case of the density. The peak velocity of the $d=800$ $\rm {\mu m}$ case is approximately 60$\%$ higher than the $d=0$ case. However the corresponding Mach number is lower because of higher ion temperature and higher sound speed. Fig.~\ref{fig:fig4} shows that we can produce a much larger dynamic range of density, temperature and velocity in a laser-driven outflow, by varying the radius of the hollow laser system. This allows a much more versatile and flexible platform for all sorts of potential laboratory astrophysics experiments, from collisionless shocks, to shear flows, jet propagation and interaction with ambient media.

The $d=0$ benchmark case has actually been realized in recent Omega experiments to study collisionless shocks \cite{Ross12, Park12}. There two plasma plumes were produced by irradiating ten Omega beams each on two facing planar targets. The hope was to achieve a laboratory collisionless shock from the plume collisions which can be used to study astrophysical collisionless shocks. To evaluate the viability of shock formation, three interesting length scales need to be considered. They are the ion collision mean free path between two counter-streaming flows \cite{Ross12, Park12},
\begin{equation}
\lambda_{mfp}(cm) \sim 5\times10^{-13}\frac{A_{z}^2}{Z^4}\frac{[v(cm/s)]^4}{n_z(cm^{-3})},
\end{equation}
characteristic electrostatic instability length scale
\begin{equation}
\l^{*}_{ES}(cm) \sim K_{1} \frac{v(cm/s)}{\omega_{pi}}\frac{W(eV)}{T_e(eV)},
\end{equation}
and characteristic electromagnetic instability length scale
\begin{equation}
\l^{*}_{EM}(cm) \sim K_{2} \frac{c(cm/s)}{\omega_{pi}},
\end{equation}
where $A_{z}$, $Z$ and $n_z$ are the ion mass in amu, average charge state and ion density, $v$ is flow bulk velocity before the collision with the other flow, $W(eV)=5.2\times10^{-13}A_{z}[v (cm/s)]^2$ is the kinetic energy per ion, $c(cm/s)/\omega_{pi}$ is the ion skin depth and $T_e$ is the electron temperature in $eV$. $K_{1}$ and $K_{2}$ are two poorly constrained numerical factors and we follow Ref. \cite{Ross12} to take $K_{1}=30$, $K_{2}=100$. A necessary but insufficient shock condition is that $\lambda_{mfp}$ has to be greater than both $l^*_{ES}$ and $l^*_{EM}$. These length scales, plus the ion skin depth at 4mm from the target for cases with $d=0$ and $d=800\,\rm{\mu m}$ are shown in the bottom panels of Fig.~\ref{fig:fig4}. In the zero radius case, the above condition is satisfied from $t\sim 2$ ns to $t\sim 8$ ns whereas in the 800 $\rm{\mu m}$ radius case, the condition is satisfied up to $t\sim$6.5 ns. Since both electron and ion density peaks around $t=5$ ns, this slight increase in collisionality has a minor effect since, we can still take advantage of the density and temperature enhancements brought about by the larger laser beams ring radius. On the other hand, there are two important advantages of the 800 $\rm{\mu m}$ radius case: (a) Both the electrostatic and electromagnetic length scales are now much smaller in absolute values, implying potentially thinner shock structures than the 4 mm collision distance \cite{Park12}. (b) The ion skin depth (bottom green dotted curve) is also much smaller, which gives the required plasma instabilities more room to operate (i.e. easier to form collisionless shock). In summary, the higher densities and temperatures of colliding jets driven by the $800$ $\rm{\mu m}$ radius beams are much more favorable for the formation and study of collisionless shocks. Even though the Mach number is reduced from $\sim18$ to $\sim 7$ (Fig.~\ref{fig:fig4}(f)), it is still high enough for strong shocks to form. 

\section{Discussion}
Regarding to the ``optimal'' hollow beam radius for jet production, there really is no definitive answer as it largely depends on what plasma platform one wants to create, and what astrophysical processes one intends to study, specifically. Some applications may require higher density, whereas high Mach number might be more important for others. Plus there are always constraints imposed by the experiment facility. The general trend we observed from our 2D simulations is the following: as the hollow laser beam radius is increased, the ablated plasma plumes collide on axis at a later time, leading to a better collimated, more axially condensed plasma jet, thus higher density, higher temperature and higher velocity. This is however, countered by the effect that as the laser beams get more spread, the laser intensity decreases for a given laser energy, leading to lower plasma ablation rate. Moreover, when the beams get too far away from each other, the plasma plumes would not have even collided yet when the laser is turned off. This means that on-axis collision would carry less and less energy and the resulting plasma jet structure would gradually fade away if the beam separation keeps increasing. At least for the Omega laser parameters, optimal density and temperature increases at 4 mm distance from laser target seem to be achieved for hollow beam radius of $\sim 800$ $\rm{\mu m}$. Because of the many competing factors, each laser configuration (e.g. NIF beams) must be studied separately for its optimal radius. However, our results demonstrate clearly that by varying the hollow laser beam radius, we can achieve much larger dynamic ranges for laser-driven plasma jets. At least for the collisionless shock experiment, a higher density allows the jet-jet collision to take place at a larger distance from the laser target. This increases the space and time scales to allow the shock to form and propagate, facilitating the observation and diagnostic of the experiment. Another beneficial effect is the generation of magnetic fields. Higher density and temperature means higher plasma pressure, which helps to generate stronger magnetic fields. The effects of these magnetic fields and the self-generated B fields by each laser spot (from Biermann Battery term), however, can only be addressed by full MHD simulations. This will be the focus of our next study phase. 

\section*{Acknowledgments}
We thank the careful review and valuable comments by an anonymous referee. This work was supported in part at the University of Chicago by the U.S. DOE through FWP 57789 under contract DE-AC02-06CH11357 to ANL. The FLASH code used in this work was developed in part by the U. S. DOE NNSA ASC- and NSF-supported Flash Center for Computational Science at the University of Chicago. Computing resource was provided by the Cyberinfrastructure for Computational Research funded by NSF under Grant CNS-0821727. This work was performed under the auspices of the U.S. Department of Energy by Lawrence Livermore National Laboratory under Contract No. B595752.
%

\end{document}